\documentclass[journal,12pt,onecolumn,draftclsnofoot,comsoc]{IEEEtran}
\usepackage[left=2.54cm, right=2.54cm]{geometry}
\usepackage[T1]{fontenc}

\ifCLASSINFOpdf
\else
\fi

\usepackage{setspace}
\usepackage{algcompatible}
\usepackage{algorithmicx}
\usepackage{algpseudocode}
\usepackage{algorithm}
\usepackage{makecell}
\usepackage{srcltx}
\usepackage{amsmath}
\usepackage{amssymb}
\usepackage{graphicx}
\usepackage{cite}
\usepackage{psfrag}
\usepackage{epsfig}
\usepackage{subfigure}
\usepackage{amsthm}
\usepackage{thmtools}
\usepackage{color}
\usepackage{stfloats}
\usepackage{makecell}
\usepackage{hyperref}
\usepackage{float}
\hypersetup{
  colorlinks   = true,    
  urlcolor     = blue,    
  linkcolor    = blue,    
  citecolor    = blue      
}
\usepackage{multirow}

\begin{document}
%
\title{\Large A Micro-Scale Mobile-Enabled Implantable Medical Sensor}

\author{M. Okwori, A. Behfarnia, P. Vuka,~\IEEEmembership{Student Members,~IEEE,} \\ A. Eslami, ~\IEEEmembership{Member,~IEEE}%



\thanks{The material in this paper was presented, in part, at the 17th International Conference on Nanotechnology (IEEE-NANO), 2016 \cite{vuka2017mobile}.}

\thanks{M. Okwori, A. Behfarnia, P. Vuka and A. Eslami are with the Department of Electrical Engineering and Computer Science, Wichita State University, Wichita, KS, USA ~(emails: \{mxokwori, axbehfarnia, pxvuka\}@shockers.wichita.edu, and ali.eslami@wichita.edu).}
}
\IEEEtitleabstractindextext{%
\vspace{-.7 in}
\begin{abstract}
Micro-scale implantable medical devices (IMDs) extend the immense benefits of sensors used in health management. However, their development is limited by many requirements and challenges, such as the use of safe materials, size restrictions, safe and efficient powering, and selection of suitable wireless communication technologies. Some of the proposed wireless communication technologies are the terahertz (THz) radio frequency (RF), and ultrasound. 
This paper investigates the use of {\em magnetic induction-based backscatter communication} as an alternative technology. In particular, the goal is to provide a practical design for a micro-scale IMD, referred to as a ``biomote'' here,  that can communicate with a wearable or handheld device such as a cell phone, tablet, or smart watch.
First, it is demonstrated that communication via magnetic induction can be established between a biomote and such an external reader. Then, low-power modulation and error-correction coding schemes that can be implemented in micro-scale are explored for the mote. With the aim of increasing reliability and accuracy of measurements through mass deployment of biomotes, suitable low-power media access control (MAC) schemes are proposed, and the feasibility of their implementation in micro-scale is highlighted. Next, assuming that the human body is an additive white Gaussian noise (AWGN) channel, the performance of the mote is simulated and analyzed. 
Results of this analysis asserts that a communication range of at least few centimeters is achievable with an acceptable bit error rate. 
Finally, from the analysis of the MAC schemes, the optimum number of motes to be deployed for various read delays and transmission rates is obtained.
\end{abstract}

}

\maketitle
\IEEEdisplaynontitleabstractindextext
\IEEEpeerreviewmaketitle


\section{Introduction\label{sec:Introduction}}

Miniaturization of IMDs to micro-scale, and ultimately nano-scale, holds great promises for enabling various in-vivo applications leading to early detection of cancer and heart attacks, deployment of nanorobots, and realization of smart drug delivery \cite{durairaj2012nano,bourzac2012carrying}.
 A number of fields are actively being investigated in the quest to develop smaller IMDs. These efforts include development of biocompatible nanomaterials, nano-electronic components, battery-less or external powering techniques, and suitable communication techniques. Biocompatible materials such as biopolymeric materials \cite{yates2013life} and DNA-based polymer coatings \cite{hong2018compact} have been proposed. Nano-scale electronic components such as diodes, bistable switches, and nanowires have been fabricated and characterized in laboratories \cite{haselman2010future}. Also, there are ongoing efforts towards battery-less, energy harvesting, or external powering of micro- to nano-scale devices \cite{liu2008mems,Deterre2012, mujeeb2015optical, Charthad2015mm,Ho2014}. A variety of communication techniques are also being investigated for deployment in IMDs. Three major approaches are the use of terahertz (THz) radio frequency (RF) communication \cite{jornet2011channel,akyildiz2016realizing,akyildiz2014teranets,akkari2016joint}, ultrasonic communication \cite{Peisino2013,Davilis2010,Galluccio2012challenges}, and optical communication \cite{schuettler2012hermetic,nafari2015metallic,song2012signal,shinagawa2013compact}. 

THz communication employs electromagnetic waves in a terahertz frequency band to communicate data, thereby resulting in two major benefits. First, it requires an antenna size of a few micrometers, which satisfies the biomote's size limitation \cite{zakrajsek2016lithographically,jornet2013graphene,llatser2012graphene}. Second, it can potentially
provide a large bandwidth. However, THz communication for biomotes undergoes extremely high attenuation in body tissues, as evidenced by a penetration depth of only up to a few hundred micrometers in fresh tissue \cite{pickwell2006biomedical,oh2013measurement}. This high attenuation limits its effective communication range inside the body to less than 1 or 2 cm. Optical and near-infrared communications also suffer from the same issues \cite{mujeeb2015optical}. 

Acoustic waves propagate very well in water and appear suitable for communication in the human body which is comprised of 65\% water. In addition, the Food and Drug Administration (FDA) allows an intensity of $7.2\ mW/mm^2$ for diagnostic ultrasound applications \cite{Charthad2015mm,Galluccio2012challenges,Davilis2010}, which is about two orders of magnitude higher than the safe RF exposure limit in the body ($10-100\  \mu W/mm^2$ \cite{federal1996guidelines,ieee1992ieee}). Despite these advantages, ultrasonic communication needs to overcome major challenges before being employed in a biomote. The large impedance mismatch at the air-tissue interface causes more than 99.95\% of the acoustic energy to be reflected back into the tissue \cite{freitas1999nanomedicine,dove2003notes,cannata2003design}. This results in a large attenuation due to the large reflection at the air-tissue interface. Also, the small size of the biomote requires a correspondingly small ultrasonic transducers to generate and detect acoustic waves. The thickness of this transducer, which is often a piezoelectric device, is inversely related to its resonating frequency. As such, the small size of the device results in a large resonating frequency. This is a significant drawback since attenuation of acoustic waves inside the body increases exponentially with both frequency and distance. For instance, at 15.4 MHz, the signal attenuation in 10 cm in blood (best case), liver, and heart (worst case) can be calculated as 31.4 dB, 115.19 dB, and 162.68 dB, respectively \cite{Galluccio2012challenges,Peisino2013}. 

 The challenges mentioned above might be overcome if a network of biomotes is deployed. \cite{pierobon2014routing,akyildiz2014teranets,akyildiz2010electromagnetic,Seo2013}. However, that bears its own complications. For THz communication, the line-of-sight requirement, and complicated implementation circuitry make implementation of the networking very difficult in micro-devices working in an uncontrolled environment such as the human body. Also, an ultrasonic-based network of motes will require one or more interrogators to be implanted under the skin in order to receive the ultrasonic signal and convert it to a radio signal that can be detected by an external wireless reader \cite{Seo2013}. 

\begin{figure}
\centering
{\includegraphics[width =3in , height=2 in]{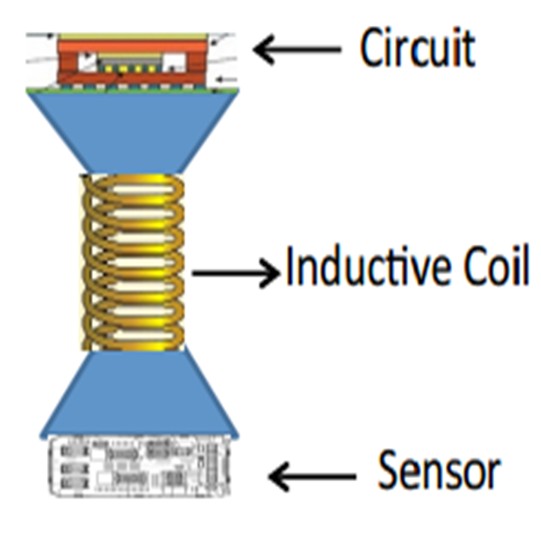}}
\caption{Biomote structure showing circuit, inductive coil and sensor.}
\label{fig:Biomote_design}
\end{figure}

\begin{figure}
\centering
{\includegraphics[width =3 in , height=2.4 in]{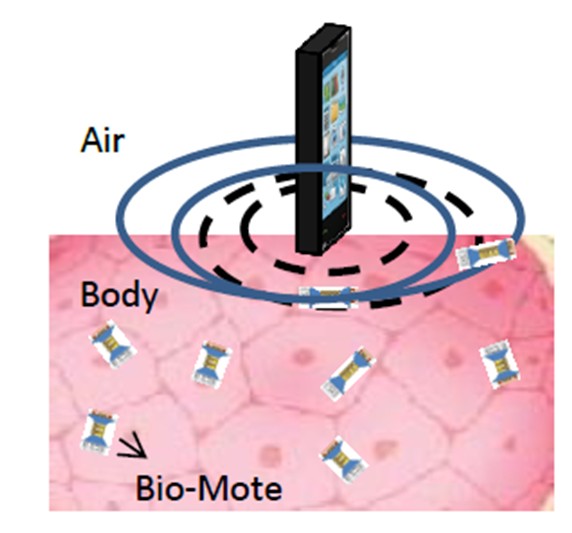}}
\caption{Biomote sending sensed data to a coupled reader.}
\label{fig:Biomote_application}
\end{figure}

An alternative means of communication is the use of magnetic fields. Magnetic induction for powering and near-field communication (NFC) with medical implants has been studied in the literature (e.g., see\cite{ghovanloo2007wide,sutardja2017isolator, Goodarzy2015,Simard2010}). While measurement ranges of a few centimeters have been achieved, the smallest implants are at least a few millimeters in each dimension. To the best of our knowledge, this work is the first comprehensive study on magnetic-induction communication and powering of implantable {\em micro}-devices. 
In this work, we present a proof-of-concept for a biomote that is externally powered and communicates using magnetic induction to a hand-held device for processing and diagnosis.
A possible design for the biomote, and how it can be used are illustrated in Figs. \ref{fig:Biomote_design} and \ref{fig:Biomote_application}, respectively.  The biomote has a communication circuitry, a dumbbell substrate, an inductive coil, and a micro- (or nano-) biosensor. When an NFC-enabled reading device, such as a cell phone, is held over the mote, the two coils are coupled, and power is induced in the mote. The mote then uses backscatter communication to send its measured data to the reading device.

The main contributions of this work are summarized as follows:
\begin{itemize}
\item We model the magnetic induction-based communication for a micro-scale coil in the mote, and provide a detailed analysis demonstrating that communication can very well be achieved within the reception sensitivity of today's handheld devices. In our analysis, we illustrate the trade-offs that emerge from the selection of resonance and subcarrier frequencies, and permeability of the coil's core material. We further discuss the impact of all the above on the communication bandwidth and signal attenuation.  
\item We consider the size and complexity limitations of the mote to choose simple and efficient modulation and error-control coding techniques. We then provide evidence of the feasibility of implementation of the selected schemes in micro-scale.
\item We evaluate the performance of the proposed physical layer, i.e., modulation and error-control coding schemes, by simulating the bit error rate at the reader. From the evaluation, we obtain the maximum distance between the mote and the reader for an acceptable BER. 
\item We then use the results of the physical layer analysis to investigate the implementation requirements and performance of simple and efficient MAC schemes. We consider three schemes: binary tree protocol, slotted ALOHA, and code division multiple access (CDMA). We also provide evidence of the feasibility of their implementation in micro-scale. In our analysis, we investigate the relationship between the mote transmission rate and the read-time required for various numbers of motes deployed. In case of CDMA, we also explore the relationship between the spreading code length and the number of successful readings from the motes deployed.  From the simulation results, we recommend optimum mote design parameters and utilization specifications.
\end{itemize}

The rest of the paper is organized as follows. Section \ref{sec:Methodology} provides a detailed design of the biomote's physical and MAC layers, including the inductive coil and selection of the modulation, error-control coding, and MAC schemes. Section \ref{sec:Results-and-Discussion} presents simulation results for the analysis of both the physical layer and the MAC layer schemes. Section \ref{sec:Conclusion} concludes the paper.

\section{Design of Micro-Scale Biomote} \label{sec:Methodology}
\begin{figure} 
\centering
{\includegraphics[width =5 in , height=2 in]{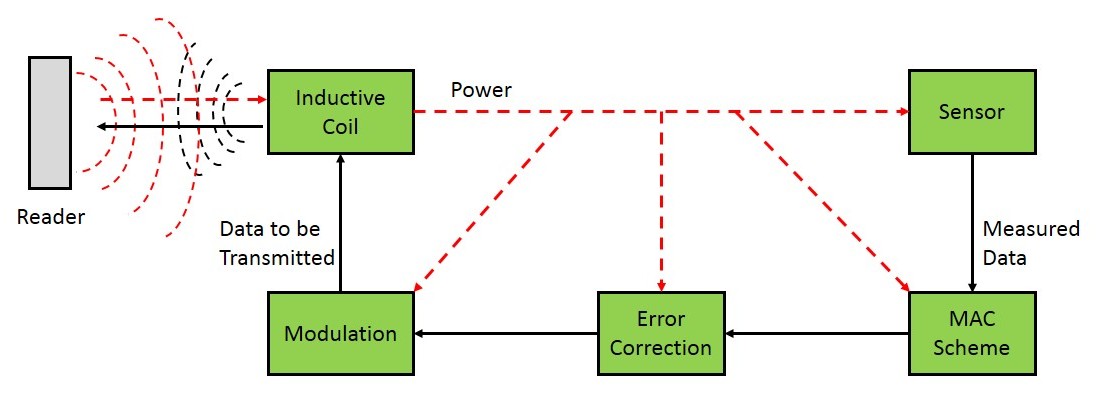}}
\caption{ Block diagram of the mote showing powering and communication systems.}
\label{fig:block_diagram}
\end{figure}

A block diagram demonstrating the building components of our proposed device is shown in Fig. \ref{fig:block_diagram}. The main blocks are the following: an inductive coil as a transceiver antenna, modulation and error-correction blocks, sensors to collect data, and a MAC scheme. Once the reader and the biomote are placed close to each other, the biomote starts collecting energy from the reader through inductive coupling, such that the induced voltage at the mote turns on all the blocks. In the following sections, the implementation of these blocks, except the sensor, is comprehensively investigated to address the major challenges for the proposed micro-scale device.



\subsection{Micro-Scale Magnetic Induction-Based Communication}\label{subsec:System-Design-and}

Environment, size of antenna, and complexity of a device are crucial points in determining the type of modulation in wireless communication. Taking these into account, and inspired by passive radio-frequency identification (RFID), we employ load modulation (a.k.a. modulated backscatter) to transmit data from the biomote to the reader. In load modulation, the impact of a coupled biomote on the reader can be modeled by a transformed impedance \cite{sun2010magnetic}, denoted here by $Z_{br}$. Controlling $Z_{br}$ at the mote using an on-off switching circuit could change the voltage at the reader. If timing of the switching circuit is controlled by data at the biomote (sensor output), this data can be transferred to the reader. The reader then applies a bandpass filter to extract transmitted data. Load modulation works in a ``near field'' where the distance between a transmitting coil and a receiving coil does not exceed $\lambda/2\pi$, where $\lambda$ is the wavelength. Since the resonance frequencies considered in this paper are less than $100$ MHz and the communication range is within a few centimeters, load modulation can be safely employed. 

\begin{figure} 
\centering
{\includegraphics[width =6 in , height=2 in]{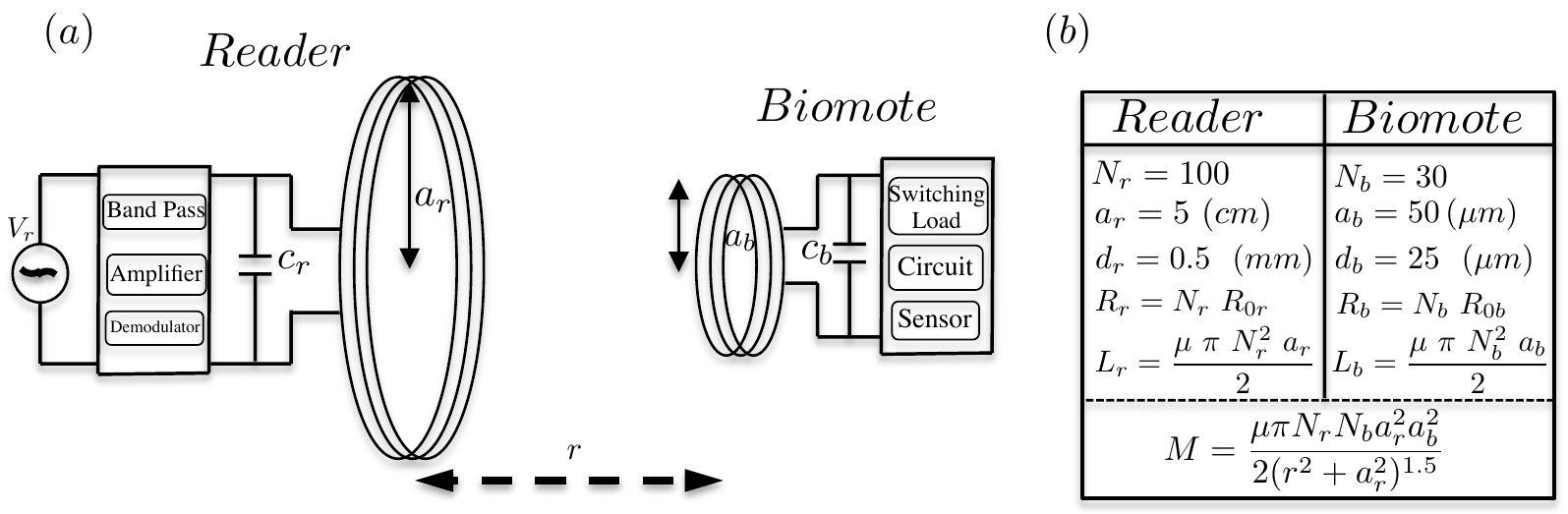}}
\caption{ Communication using magnetic induction between reader and biomote: (a) structure, (b) parameters of equivalent circuit: $N$ is the number of turns, $a$ is the radius of a coil, $d$ is the diameter of wire (thickness), $r$ is the distance between devices, $R_0$  is the resistance of unit length, $R$ is the resistance of a coil, $L$ is self inductance, and $M$ is mutual induction between coils.}
\label{fig:C_2_C}
\end{figure}

In order to measure the received power at the reader, we use equivalent circuits for transceiver coils. Fig. \ref{fig:C_2_C} shows the structure and parameters of the equivalent circuit of antenna coils. We select the parameters in a way that the hight and diameter of the biomote does not exceed $250 \,  \mu m$.
Here, we use copper and gold for the coils of the reader and biomotes, respectively. Also, we assume that the voltage in the reader is $3.8$ v (the voltage of iphone 6 \cite{wiki_phone}).
As can be seen in Fig. \ref{fig:C_2_C}, coils on both sides along with parallel capacitors ($C_r$ and $C_b$) form resonant circuits. For our analysis in this paper, we consider three resonant frequencies: $1$, $13.56$, and $100$ MHz. It should be noted that, due to the skin effect, this frequency affects the resistance of the coils, and hence, their impedance.  
The power transmitted by the reader and received at the biomote can be obtained as
\begin{align}
\notag &P_t = Re \Big\{ \frac{ V_{r}^2}{( Z_{br} + Z_r )} \Big\}, \\
&P_r = Re \Big\{ \frac{Z_L \ V_{rb}^2}{( Z_{rb} + Z_b + Z_L)^2} \Big\},
\label{eq:power_r}
\end{align}
respectively. Her, $Z_b$ is the impedance of the coil at the biomote, $Z_r$ is the impedance of the coil at the reader, $Z_{br}$ is the impact of biomote 's impedance on the reader, $Z_{rb}$ is the impact of reader's impedance on the biomote, $V_{rb}$ is the induced voltage by the reader on the biomote, and $Z_L$ is the impedance of the load at the biomote. The analysis leading to eq. (\ref{eq:power_r}) is well-known, and its details can be found in \cite{sun2010magnetic}. 
It is worth noting that these equations hold for the best case scenario when the two coils are positioned in a coaxial fashion. However, the analysis can be extended to a biomote with a tri-directional (3D) coil to address the problem of orientation. More information about 3D coils can be found in \cite{3DCoil97} and \cite{3DCoil14}.

\begin{figure} [t]
\centering
{\includegraphics[width =4.6 in , height=3.2 in]{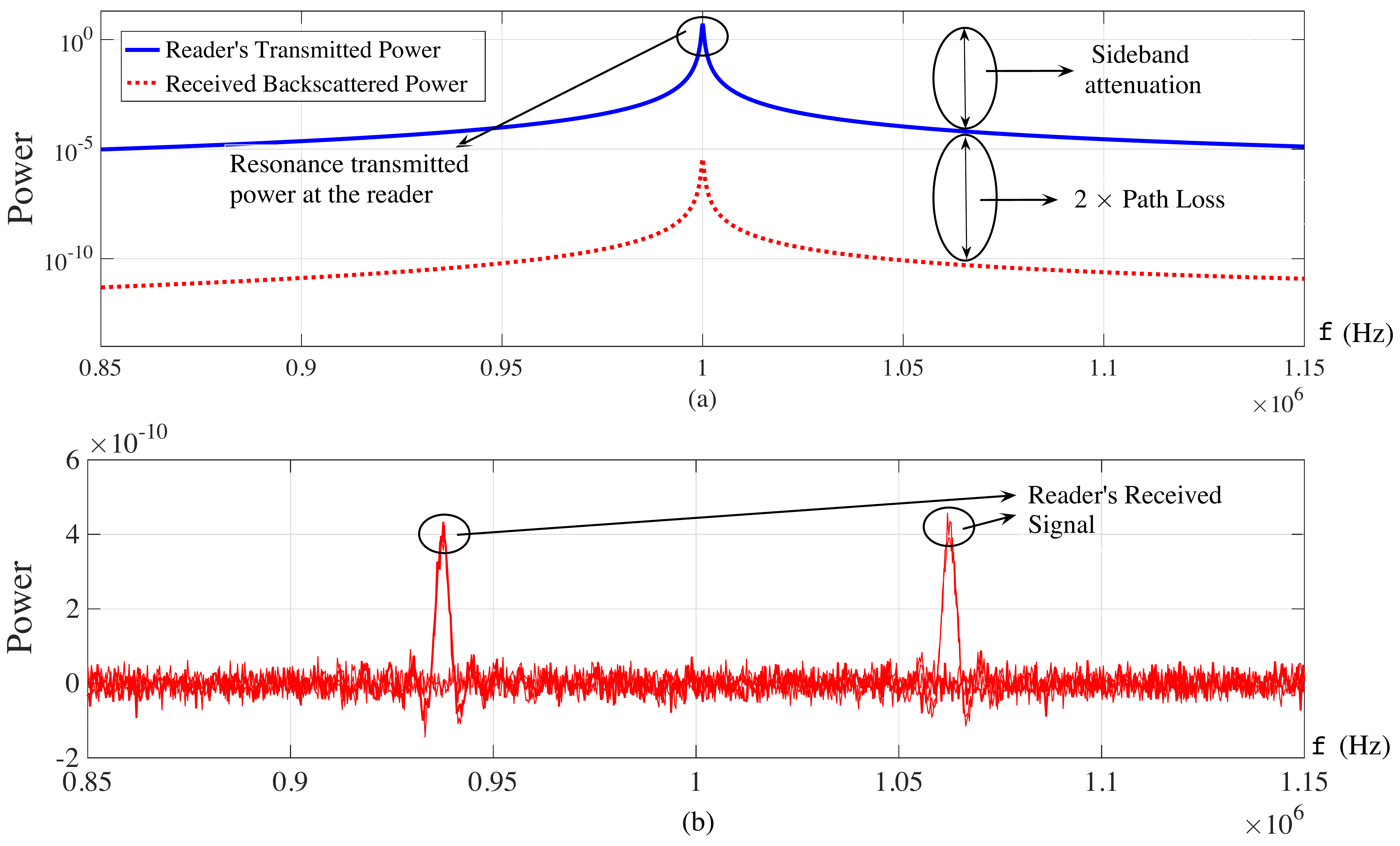}}
\caption{ Signal powers at the reader: (a) transmitted and received powers at the reader, (b) received backscatter power at the reader compared to noise. Parameters are chosen as follows: $d= 6$ cm (distance between devices), $f_c = 1$ MHz (resonance frequency), and $f_s = 62.5$ KHz (subcarrier frequency).
}
\label{fig:BackS}
\end{figure}

To gain a better understanding, Fig. \ref{fig:BackS} shows both the transmitted power and the received power at the reader. As can be seen, the transmitted power peaks at the resonance frequency ($f_c$),  $1$ MHz in this figure. The received backscatter power (from the biomote) experiences a roundtrip path loss ($PL$) and a sideband attenuation ($A_s$). To minimize $PL$, regardless of changing the coils' parameters such as the number of turns, one could use ferrite cores with high permeability materials in both of them. To reduce $A_s$, as shown in Fig. \ref{fig:BackS}, we can see that the closer the sideband is to the resonance frequency, the less we are affected by the attenuation. However, if the distance between the resonance frequency and the subcarrier frequency ($f_s$) is reduced, the available bandwidth will also be reduced. Hence, the trade-off between attenuation and bandwidth should be considered in an optimum design. 
It is also worth mentioning that the subcarrier signal at the mote can be obtained from the carrier signal using a frequency divider which is essentially a very simple shift register. Therefore, we often have $f_s = \frac{f_c}{2^n}$, where $n$ is the length of the shift register.

 \renewcommand{\arraystretch}{1.6}
\begin{table}
\centering
\caption{Received backscatter power at the reader from the biomote. Distance between the two devices is $6$ cm. Here, $P_{tx}$ and $P_{re}$ are the peaks of the transmitted power and the received backscatter power at the reader, respectively. $f_s$ is the frequency of switching load at the biomote, $L$ is path loss, and $A_s$ is sideband attenuation.}
\label{tb:calc_fc}
\begin{tabular}{|c|c|c|c|c|c|c|c|}
\hline
\multicolumn{1}{|p{4.8em}|} {\footnotesize Resonance Frequency (MHz)} & $R_r$ \footnotesize $(\Omega)$  & $P_{tx}$ \footnotesize $(dB)$ & $R_b$ \footnotesize $(\Omega)$ & $ \mu$ & $f_s$ &$P_{re}$ \footnotesize $(dBm)$ & \multicolumn{1}{p{7.3em}|} {\small Total Loss \footnotesize$(dB)$ \quad \ \ \ $=\ 2\, (PL) + A_s$} \\
\hline
 \renewcommand{\arraystretch}{1.55}
$f_c = 1$ & $5.881$ & $3.90$ & $0.7533$ &  \makecell{$1$\\ $10$ \\ $50$}  & \Large $\frac{f_c}{16}$ & \makecell{$- 128.37$ \\ $-108.38$  \\ $-95.47$} &  \makecell{ $2\, (61.14) + 40$ \\  $2\,(41.14)+60$  \\  $2\,(27.17)+75$ }  \\ \hline
$f_c = 13.56$ & $19.64$ & $-1.34$ & $0.7542$ & \makecell{ $1$ \\ $10$ }  & \Large  $\frac{f_c}{64}$ &  \makecell{$-98.82$ \\ $-80.88$ } &  \makecell{ $2\,(43.74)+40$ \\  $2\,(23.77)+62 $}  \\  \hline
$f_c = 100$& $52.13$ & $-5.58$ & $0.8577$ & $1$ & \large  $\frac{f_c}{128}$& $- 87.95$ &  $2\,(31.18)+50$ \\
\hline
\end{tabular}
\end{table}

Table \ref{tb:calc_fc} shows the received backscatter power at the reader which is located $6$ cm away from a biomote. The calculations are obtained for different values of $f_c$ and relative permeability ($\mu$) of the ferrite core. As can be seen in Table \ref{tb:calc_fc}, increasing $f_c$ reduces path loss. In addition, since $f_c$ is directly proportional to $f_s$ as explained above, increasing $f_c$ leads to a larger bandwidth. 
For instance, if $f_c$ is increased from $1$ MHz to $13.56$ MHz, then the bandwidth will also increase from $120$ KHz to $1690$ KHz, assuming that $f_s = f_c/16$.
This achievement, however, is obtained at the price of a higher quality factor ($Q$), which implies higher sideband attenuation. Therefore, increasing $f_c$ reveals a trade-off between available bandwidth and total loss. To address this, one could reduce $f_s$ to avoid severe sideband attenuation, while maintaining an acceptable bandwidth. For example, in Table \ref{tb:calc_fc}, when $f_c$ is increased from $1$ MHz to $13.56$ MHz, $f_s$ is reduced four times.
This reduction keeps sideband attenuation low, while providing approximately $400$ KHz of bandwidth which can be considered acceptable given the bit rate requirements of our particular application.

Another factor that impacts received power is relative permeability or $\mu$. The $\mu$ of a medium that mostly contains air and blood is expected to be around $1$. Although the permeability of a medium is not in our control, one could fill the core of a coil and its surrounding package with a high permeability material such as ferrite to increase the received power. The permeability of ferrite, depending on a number of factors, can be as high as $1700$ \cite{hu2005science}. Since the medium here is a combination of the mote, the human tissue, and the air, its permeability will be a function of that of the ferrite core, the human body, and the air. The core, with the highest permeability, is only a fraction of the medium (approximately $1/240$). Taking this into consideration, the resultant permeability of the entire medium is assumed to be about $10$.
 
We would like to make two more points regarding the use of highly permeable cores, which also help in further understanding the numbers in Table \ref{tb:calc_fc}.  Although higher values of $\mu$ decrease path loss, they cause higher sideband attenuation. Hence, total loss needs to be paid attention to while considering the use of a high permeability core. Another drawback of high permeability materials is that their $\mu$ often decreases with increasing frequency. The value of $\mu$ for ferrite, for example, could reduce from more than $600$ to $1$ as the frequency increases from $1$ MHz to $100$ MHz \cite{hu2005science}. Thus, while increasing the frequency provides a higher bandwidth, it practically negates the effect of $\mu$ on reducing $PL$. 


The sensitivity of the reader also plays a vital role in detecting the backscattered power from the biomote. Sensitivity is the lowest signal power from which  useful information can be obtained. Given the sensitivity, usable received signals are usually within certain power levels. For example, the reference signal receive power (RSRP) as the LTE signal strength indicator could range from $-45$ dBm for excellent signals to $-140$ dBm for poor signals \cite{LTE_RSRP}\cite{LTE_RSRP2}. Taking $-100$ dBm as sufficient received power at the reader, design parameters could be selected similar to those of Table \ref{tb:calc_fc}. Section \ref{sec:Results-and-Discussion} provides more on the achievable communication range and quality as a function of design parameters and noise power.

\subsection{Modulation and Error-Correction Coding for Micro-Scale Design}\label{subsec:physical_layer_design}

Taking complexity and power consumption into consideration, ASK and BPSK are selected for load modulation in the biomote. These modulation techniques are among the simplest digital modulation techniques, and are employed in various biomedical implants\cite{cho2006ask}. Their implementation has also been demonstrated in micro-scale \cite{Hassouni2015design,burasa2016high}, with rates of $160$ kbps and $500$ kbps reported, respectively.

In order to increase the efficiency of magnetic induction-based communication, error-correction coding is introduced at the biomote. Though complex coding schemes can achieve low bit error rates, they are computationally intensive and, hence, unsuitable for the biomote with its size and power constraints. Short linear block codes, namely Hamming (15,11) and Reed-Solomon (31,26), were selected  since they are less complex and can be implemented in micro-scale. Both codes can be implemented on a very small die area using shift registers \cite{yang2012product}.
In this work, since the biomote can only send information, it will need to only implement the encoding circuit. Decoding will be handled at the receiver with much more power and high-end processor. This is quite advantageous because in most error-correction mechanisms, decoding is a significantly more complicated process than encoding.

Hamming code (15,11) has a minimum distance of four, thus can detect three errors and correct one error. Reed-Solomon (RS) codes perform better than Hamming codes because they have the ability to correct both random and burst errors. RS codes are constructed using generator polynomials of degree $2t$. An RS (n,k) code can correct $t$ errors, where $t=\lfloor \frac{n-k}{2} \rfloor$. The RS codes (31,26) can, therefore, correct two errors. Hamming (15,11) provides a code rate of 0.733, compared to 0.8387 for RS (31,26). 

\subsection{MAC Layer for Micro-Scale Design}\label{subsec:mac_layer_design}
The probability of failure of micro-scale sensors warrants the deployment of a large number to capture readings that can be averaged to obtain more accurate and realistic measurements. MAC layer protocols are very well advanced and widely deployed, but their implementation in micro-scale needs to be investigated and analyzed. In this section, we evaluate lightweight MAC schemes, identify their implementation requirements, and investigate the feasibility of their implementation in micro-scale.

The particular scheme utilized depends on factors such as the complexity of nodes, performance requirements, and the network structure. 
Due to the power and complexity restrictions at the mote, we consider three MAC schemes as candidates for implementation at the mote, including two dynamic protocols and one static scheme. The three candidates are evaluated for simplicity and efficiency, and two are selected for simulation and analysis.  The schemes considered are binary tree protocol, slotted ALOHA, and CDMA, each of which has specific design requirements. Table \ref{tb:mac_implementation} summarizes the required components in each scheme and provides evidence of their implementation in micro-scale in the literature.
As can be noted, we ruled out TDMA and FDMA due to the additional complexity imposed on mote's circuitry. FDMA requires implementation of an oscillator and other circuitry on the mote to generate the sub-carrier frequency \cite{bletsas2009anti}. TDMA requires a network controller to assign unique time slots to different motes, via extra messages exchanged between the motes and the reader  \cite{das2014performance}.
 
\renewcommand{\arraystretch}{1.6}
\begin{table} [t]
\small
\centering
\caption{Requirements and Evidence of Feasibility of Implementation of Considered MAC Schemes. }
\label{tb:mac_implementation}
\begin{tabular}{|c|c|c|}
\hline
\multicolumn{1}{|p{1em}|}{ Scheme} & { Requirements} & \multicolumn{1}{p{14em}|}{  Implementation in Micro-Scale}\\ \hline

Binary Trees &  \makecell{Unique Address\\ Synchronization \\ Internal Clocks}  &  \makecell{Shift Registers \cite{yang2012product} \\ J-K Flip-Flop Synchronization Circuit \cite{salles2017dissipated} \\ Ferromagnetic
Cladding \cite{zhang2014low}}\\ \hline

Slotted ALOHA & \makecell{Unique Address\\ Synchronization \\ Internal Clocks \\ Random Number Generators}     &  \makecell{Shift Registers \cite{yang2012product} \\J-K Flip-Flop Synchronization Circuit  \cite{salles2017dissipated} \\Ferromagnetic
Cladding \cite{zhang2014low}\\ Feedback Shift Registers \cite{Datta2017}}   \\ \hline
 
CDMA &  \makecell{Spreading Codes \\ Synchronization \\ Internal Clocks \\ Chipset and Data Multiplier}       &   \makecell{Pre-Generated and Stored on Shift Registers \cite{yang2012product}  \\J-K Flip-Flop Synchronization Circuit  \cite{salles2017dissipated} \\Ferromagnetic
Cladding \cite{zhang2014low}\\  XOR Gates \cite{bharti2017implementation}}       \\

\hline
\end{tabular}
\end{table}

The binary tree protocol is used to select one mote from those in the interrogation zone of the reader. This is done by the reader sending commands and narrowing it down to a particular mote. The average number of iterations required to select a single mote from $N$ motes is given as 
\begin{equation}
L(N)=\frac{\log{(N)}}{\log{2}}+1.
\end{equation}
Each mote is given a unique address and is required to synchronize its response to identification commands with all other motes. Synchronization among sensors has been tackled in a number of ways in the literature. The scheme used depends on the desired precision, duration of synchronization, area of sensor network, whether all or just portions of nodes need to be synchronized, energy, and infrastructure constraints in achieving the synchronization scheme \cite{Cox2005time}. In our design, we utilize the unidirectional scheme, where the reader sends a syncing message to all motes after powering. This resets and synchronizes the individual clocks of the motes. An appropriate sequence detection circuit using J-K flip-flops has been designed in micro-scale in \cite{salles2017dissipated}. The high number of exchanges of command messages required to select a mote makes this scheme less efficient than the other two schemes for our design. Therefore, it is not included in the simulation results presented in the next section. 

A simple MAC scheme is slotted ALOHA. This scheme is reader driven and stochastic in scheduling the reception of packets from motes. It utilizes a randomly generated wait time in the cyclic sending of data, which ensures a fair probability that more than one mote can successfully transfer readings without collision. 
Implementation of the slotted ALOHA scheme requires time synchronization in all motes and  random number generators. In the literature, a couple of suitable lightweight techniques have been proposed to generate pseudo-random numbers \cite{Luby1996}. We propose to use a linear feedback shift register (LFSR)-based circuit because it provides high-speed binary sequences and can be miniaturized \cite{Datta2017}. The authors of \cite{mandal2016} report implementation of a pseudo-random number generator based on a nonlinear feedback shift register using a micro-scale CMOS process.

Another option we consider is CDMA, which involves multiplying the data with orthogonal spreading codes before modulation and transmission. The receiver is able to separate and decode data from each sensor by correlating the received signal with the code of all sensors sequentially. Implementation requires the synchronization of transmission times from all sensors because non-synchronization results in multi-access interference \cite{benkic2007proposed}. The problem of good cross correlation between the spreading codes to prevent primary collision is limited in our application because the number of motes contending for access at a point in time is limited. For implementation in micro-scale, XOR gates can be used to multiply the spreading code with the sensed data, and the static spreading codes are assigned to the motes during the design stage and stored using registers. Implementation of XOR gates in micro-scale has been demonstrated in \cite{bharti2017implementation}. 

\section{Simulation Results}\label{sec:Results-and-Discussion}
To bring clarity and make concrete sense of the proposed design for both physical layer and MAC layers, we set up numerical simulations for several realistic scenarios. Through physical-layer evaluation, we derive BERs corresponding to error-correction coding techniques described in section \ref{subsec:physical_layer_design} for various communication ranges between the reader and biomotes. Taking into account the results obtained in the physical-layer analysis, particularly the communication range, we evaluate the MAC protocols described in section \ref{subsec:mac_layer_design}. The analysis provides useful insights to questions such as: What read-time is required for a given data rate and number of deployed motes at a specific location of interest? What data rate should be employed at the mote for a desired read duration and a total number of motes deployed? From simulation results, design parameters and utilization specifications can be recommended to answer these questions for different scenarios. 

\subsection{Physical-Layer Simulation Setup}\label{sec:sim-setup}
The focus of the physical-layer analysis is on determining the maximum distance between the mote and the reader for which communication can be achieved. The magnetic induction-based model is simulated using the design parameters shown in Table \ref{tb:parameters_sim}. 
In this design, we assume that noise power is the summation of thermal noise and noise figure. Thermal noise is expected to be around $-120$ dBm for a bandwidth of $200$ KHz and temperature of $290^{\circ}$ \cite{wiki_noise}. Also, the noise figure, which is added by electronic circuits is expected to be $15$ dB or less \cite{Noise_Fig}. Taking the worst case, we assume $-105$ dBm as total noise at the reader. We also assume that the core of coils are filled with air, which implies $\mu = 1$.
The power induced in the external reader is evaluated for various separation distances. We simulate the BER achieved using ASK and BPSK for three conditions each: without error-correction coding, with Hamming (15,11) code, and with RS (31,26) code.
 \renewcommand{\arraystretch}{1}
\begin{table}
\centering
\caption{Simulation Parameters for Physical-Layer Analysis.}
\label{tb:parameters_sim}
\begin{tabular}{|c|c|}
\hline
 Parameters & Value \\ \hline
 External Coil Radius & $5$ cm \\
Implant Coil Radius & $50$  $\mu$m \\
Noise Power &  $-105$ dBm \\
Carrier Frequency & $13.56$  MHz \\
Permeability & $1$ \\
\hline
\end{tabular}
\end{table}

\subsection{Physical-Layer Simulation Results}\label{sec:sim-setup}
\begin{figure}
\centering
{\includegraphics[width =4.8 in , height=2.8 in]{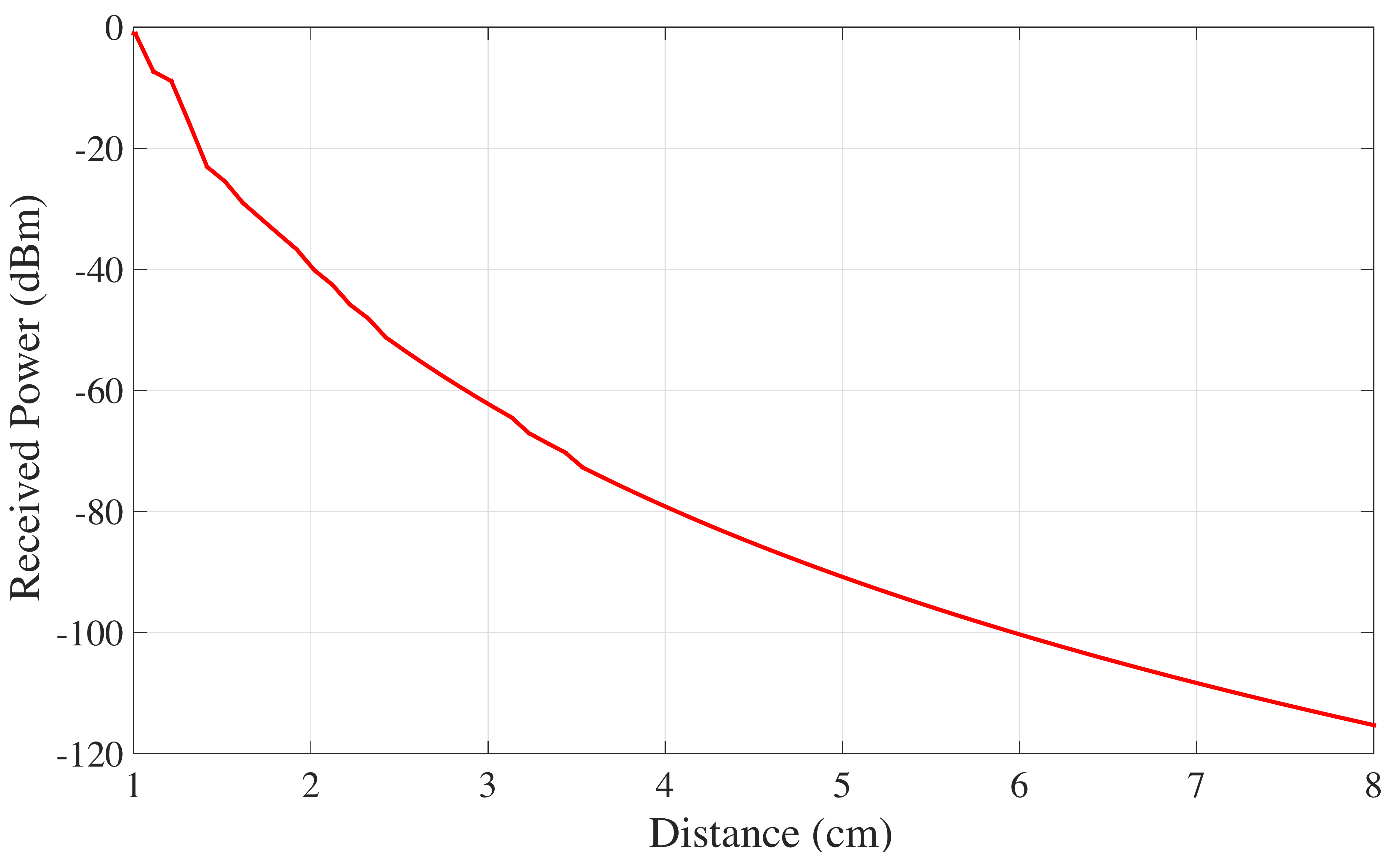}}
\caption{ Backscattered induced power in external reader from biomote as a function of separating distance between two devices.}
\label{fig:power}
\end{figure}

Fig. \ref{fig:power} shows the magnitude of backscattered power induced in the external reader with respect to the distance between the two devices.  As the distance increases above $6$ cm, received power falls below $-100$ dBm, making the signal detection difficult. As a result, given the design parameters of Table \ref{tb:parameters_sim}, the effective communication range would be around 6 cm.

\begin{figure} 
\centering
{\includegraphics[width =4.8 in , height=2.8 in]{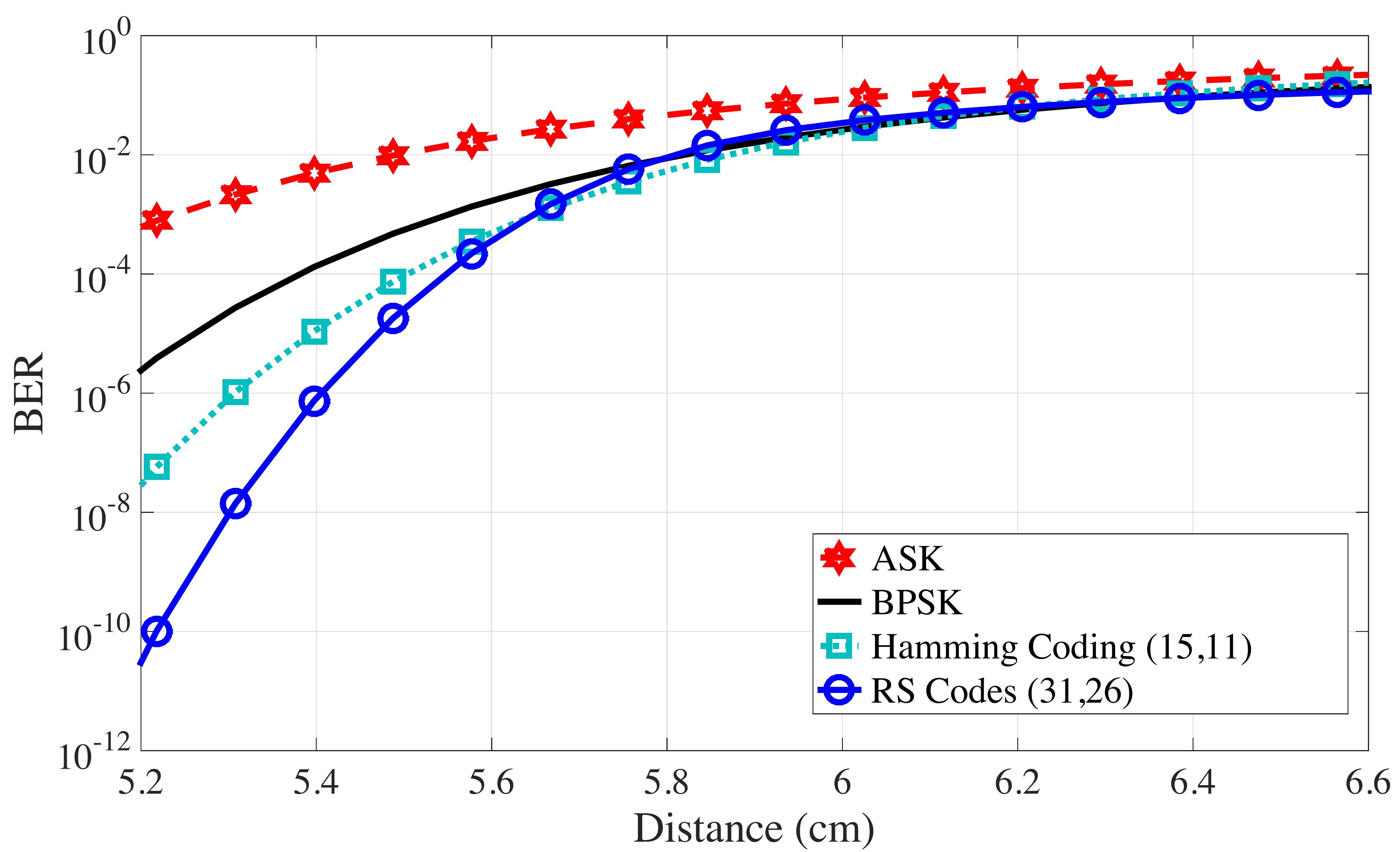}}
\caption{ Performance of using error-correction coding in the system: bit error rate (BER) versus various separating distances between reader and biomote.}
\label{fig:codings}
\end{figure}

Fig. \ref{fig:codings} depicts system performance in terms of BER for ASK, BPSK, BPSK alongside the Hamming code, and BPSK alongside the RS code. It can be observed that the BER for distances above $6$ cm reaches above $0.01$ in all cases.  
In our application, given the short packet length (see Section \ref{sec:sim-setup}) and large number of motes deployed, it is safe to assume that a BER of about $10^{-3}$ would be acceptable. According to the figure, this BER is achievable for distances between $5$ to $6$ cm.
When comparing the performance with respect to both modulation and error-correction coding, BPSK along with the RS code performs the best. In general, various implants may require different BERs, depending upon data rate and the format of data transmitted. Once the required BER is known, the respective combination of modulation and error coding can be selected according to this figure. 

\subsection{MAC Scheme Simulation Setup}\label{sec:sim-setup}

The main focus of our analysis for slotted ALOHA is to address the question of how long it would take the reader to obtain readings from all motes in its interrogation zone, and also to address the concern of what transmission rate to deploy at the motes. The three parameters of interest in our investigation are the time duration to obtain readings from all motes, transmission data rate of the motes, and number of motes contending for access at a particular location. These three parameters are henceforth referred to as $Read-time$, $Rate$, and $Number\ of \ Motes$, respectively. Two scenarios are considered in evaluating the performance of slotted ALOHA. Scenario 1 corresponds to a global deployment of the motes over the entire human body, and Scenario 2 to a local deployment of motes at only one region of interest. Simulations are carried out for packet lengths of $8$ and $64$ bytes.

In Scenario 1, the $Number\ of \ Motes$ at a particular location is not constant but rather a function of the total number of motes globally deployed, the volume of the human body, and the area of the interrogation zone of the reader. As such, for the $Number\ of \ Motes$ increasing in steps of $10$, we evaluate the $Read-time$ required to read all motes at various $Rates$. The interest here is to recommend the $Read-time$ required for the desired $Rate$, if the total number of motes deployed globally is known. For this analysis, we use the average volume of the human body and the volume of the interrogation zone of the reader. The volume of the interrogation zone is determined using a maximum communication range of $5$ cm which is recommended by our physical layer analysis. In Scenario 2, the $Number\ of\ Motes$ is fixed and is equal to the number of motes locally deployed. In this case, we set the $Read-time$ to values between $2$ and $10$ seconds in steps of $2$ seconds, and evaluate the $Number\ of \ Motes$ with successful transmissions. Our interest here is to recommend the number of motes to be deployed locally for various values of $Rate$ and $Read-time$.

In our simulation of the slotted ALOHA, we assume that each mote would have one data packet for transmission. It is also assumed that the reader is capable of processing all received packets during one slot duration. As a result, no queues are formed at either the mote or the reader. Our analysis do not consider the time taken to receive packet acknowledgments, and it is assumed that packet failures are due to collisions as a result of multiple transmissions on a slot. For both scenarios, we ran $100$ simulations to obtain the average number of successful motes out of the $Number\ of \ Motes$. 

The focus of our analysis for CDMA is to determine how many motes can successfully communicate with the reader in a reading session. We evaluate this by randomly assigning static spreading codes of lengths $16$, $32$, $64$, $128$, and $256$ to the motes, and uniformly sampling from the pool of motes. Our analysis is based on the assumption of equal transmission energy for motes and synchronization of the start time of transmissions. Also, we only consider multiple access interference due to data transmissions of other motes. 

In addition to the above, we present a performance comparison of the two MAC schemes for two read durations: short and long. The short read duration and the long read duration are arbitrarily set to $128$ slots and $128 \times 10$ slots, respectively. For the simulation, packet length is set to 64 bytes and transmission rate to 20 kbps, resulting in a short read duration of $3.28$ seconds and a long read duration of $32.8$ seconds.

\subsection{Results of MAC Layer Analysis }\label{sec:sim-setup}
\begin{figure}[t]
\centering
{\includegraphics[width =5.4 in , height=3.7 in]{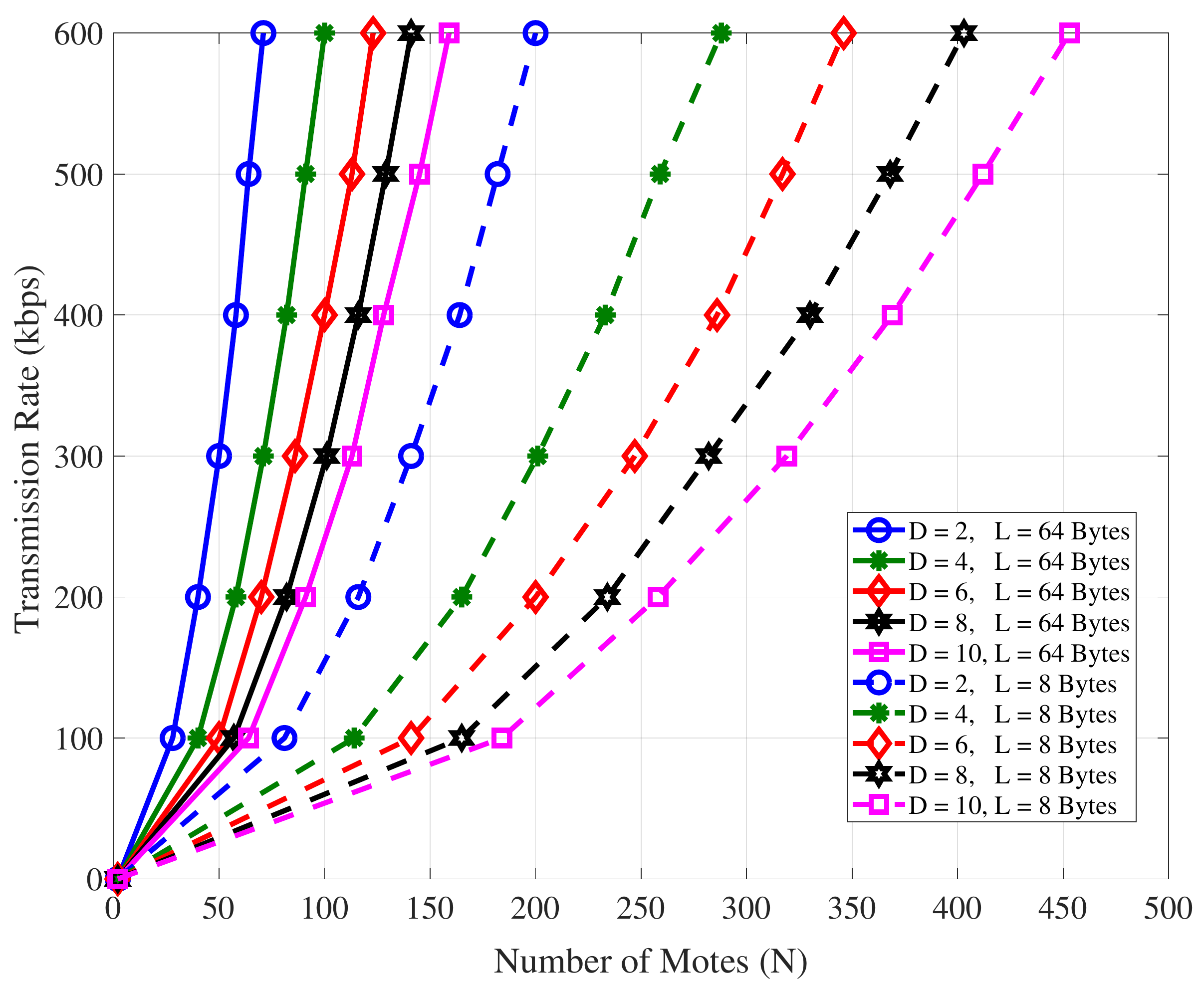}}
\caption{Transmission rates to achieve successful transmission from all motes for various numbers of contending motes in specific read-times.}
\label{fig:scenario_1}
\end{figure}

Results of the Scenario 1 analysis are shown in Fig. \ref {fig:scenario_1}. For $Rates$ varying from $1$ to $600$ kbps, we obtain the highest number of contending motes that all achieve successful transmissions in specific $Read-times$. This was done for data packet lengths of 8 and 64 bytes. From this, we can infer what combinations of $Read-time$, $Rate$, and $Number\ of \ Motes$ are feasible. For instance, for a mote operating at $200$ kbps, at a $Read-time$ of $10$ seconds and a packet length of 64 bytes, $91$ motes can be successfully read in the interrogation zone. Therefore, assuming an interrogation zone of $5$ cm radius and average volume of the human body as $6.643\times 10^5\ cm^3$,
\begin{figure} [H]
\centering
{\includegraphics[width =5.4 in , height=3.7 in]{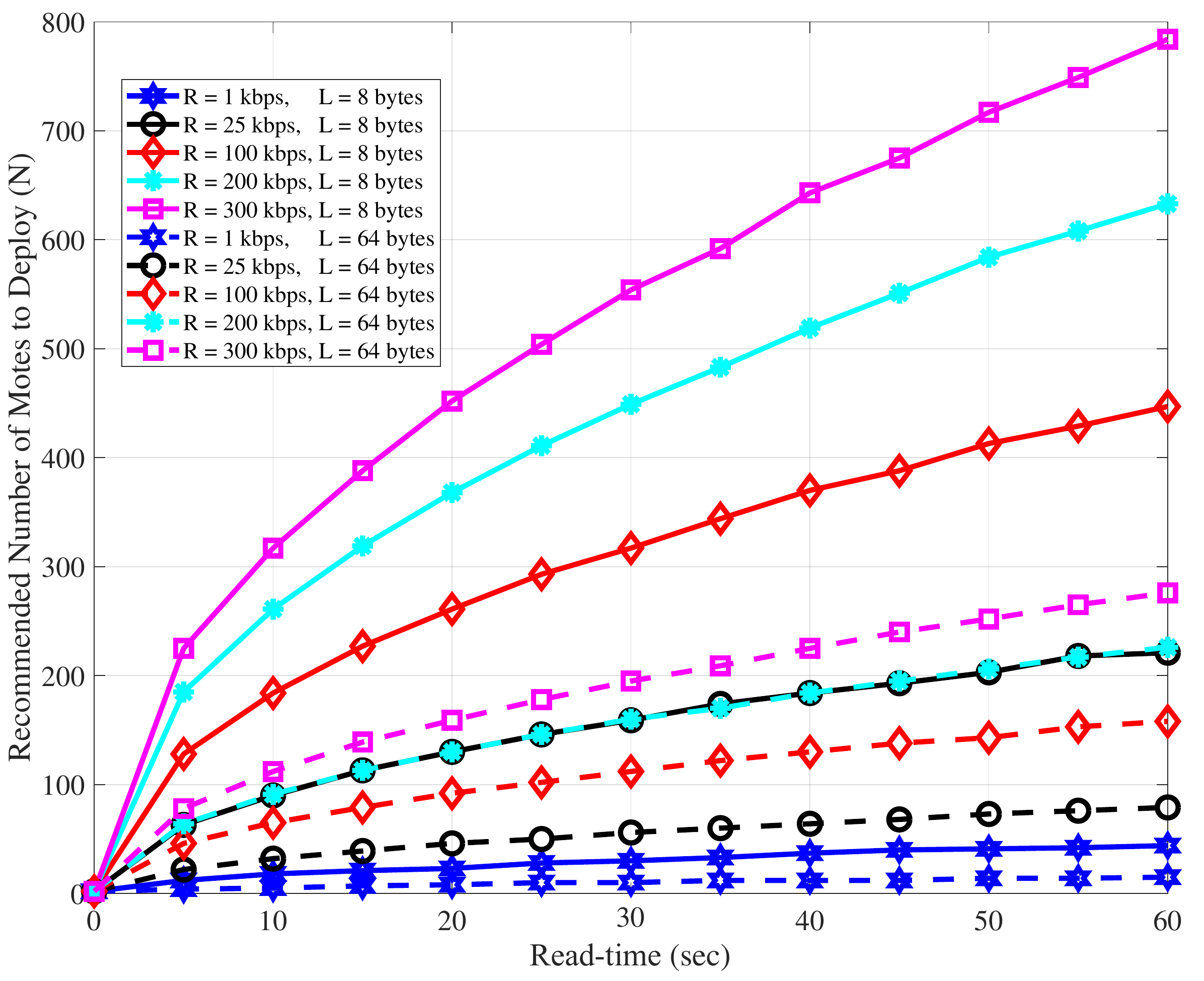}}
\caption{Recommended number of motes to locally deploy for various rates and read-times.}
\label{fig:scenario_2}
\end{figure}
 \noindent the recommended $Number\ of \ Motes$ to be deployed globally in the human body would be $230,906$. Similar deductions can be made for a data-packet length of 8 bytes. The packet length utilized depends on the application requirement. For Scenario 2, the results are shown in Fig. \ref {fig:scenario_2}. From these results, the optimum number of motes for local deployment can be recommended for any choice of $Rate$ and $Read-time$. For example, for a mote operating at a rate of $200$ kbps with 64-byte data packets, an optimum number of $130$ motes can be deployed for a desired read duration of $20$ seconds.

\begin{figure}[t]
\centering
{\includegraphics[width =3.5 in , height=2.34 in]{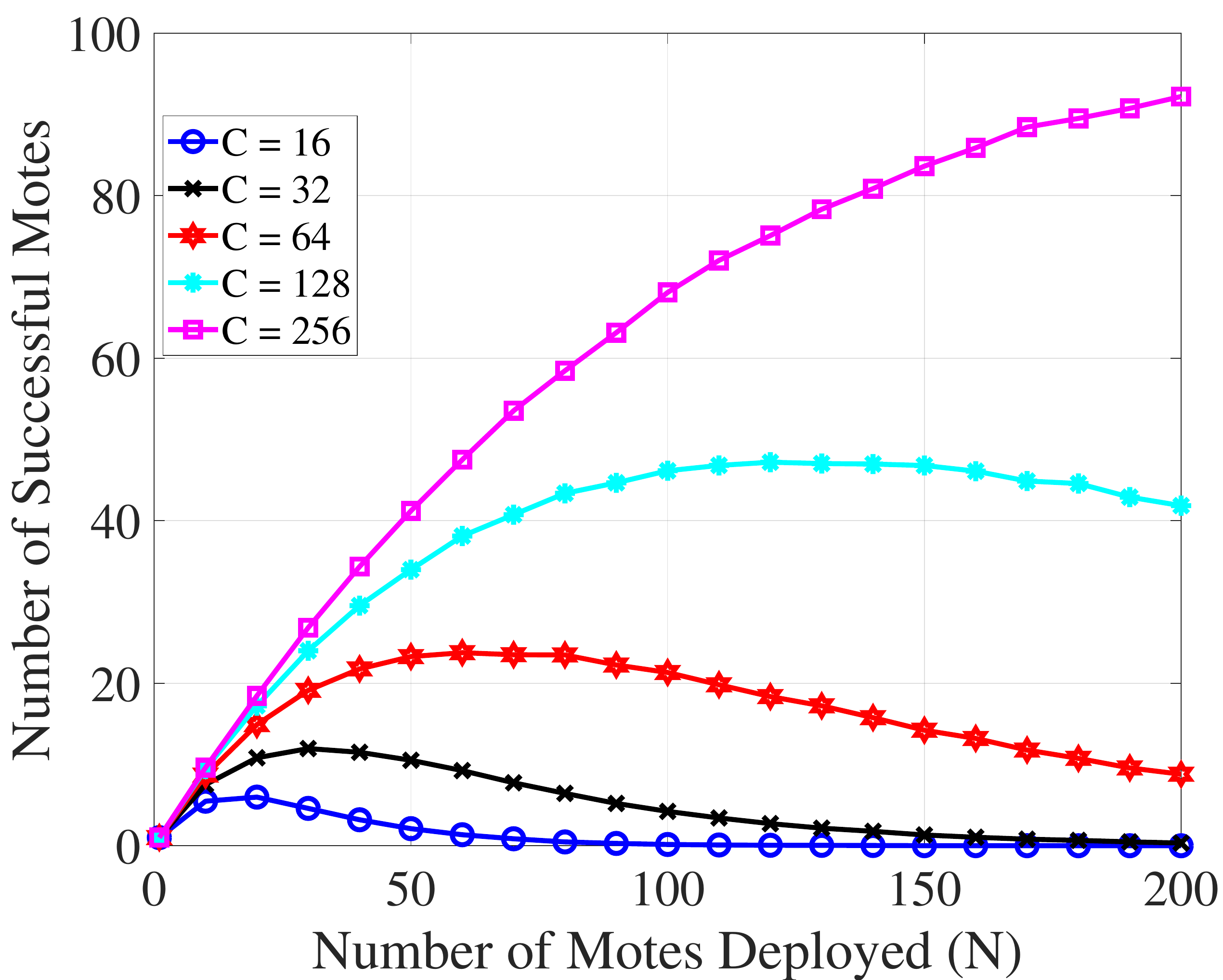}}
\vspace{-0.1 in}
\caption{Number of successful transmissions for various number of motes using static CDMA codes of length C.}
\label{fig:cdma_throughput}
\end{figure}

Results of the CDMA analysis are shown in Fig. \ref{fig:cdma_throughput}. As can be seen, there is a positive correlation between the number of successful motes and the length of codes. It was observed that for every length of codes, the number of successful motes rises to a peak and then drops as the number of motes contending for access increases beyond a particular point.

\begin{figure}[t]
\centering
{\includegraphics[width =3.5 in , height=2.34 in]{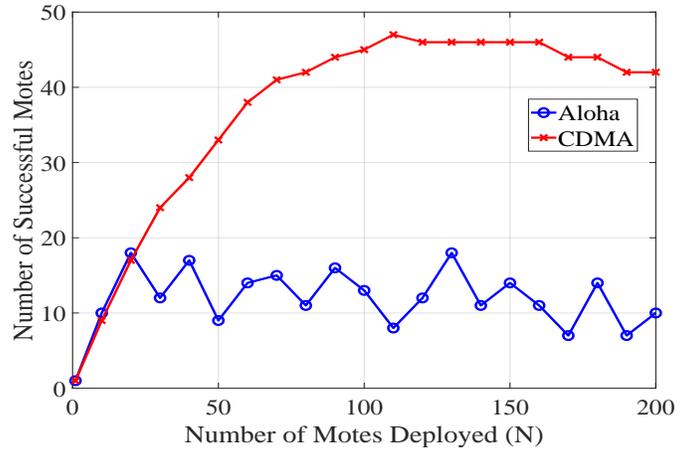}}
\vspace{-0.1 in}
\caption{Comparison of successful transmissions for CDMA and Slotted ALOHA for short-read durations, D = $128$ slots, data rate is $20$ kbps and packet length is $64$ bytes.}
\label{fig:aloha_vs_cdma}
\end{figure}

\begin{figure}[t]
\centering
{\includegraphics[width =3.5 in , height=2.34 in]{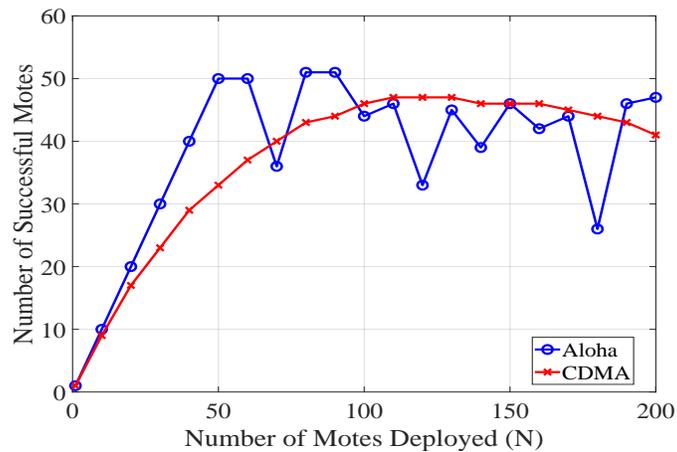}}
\vspace{-0.1 in}
\caption{Comparison of successful transmissions for CDMA and Slotted ALOHA for long-read durations D is $128 \times 10$ slots, data rate is $20$ kbps and packet length is $64$ bytes.}
\label{fig:aloha_vs_cdma2}
\end{figure}

Figs. \ref{fig:aloha_vs_cdma} and \ref{fig:aloha_vs_cdma2} show results of the comparison of the two MAC schemes. In  Fig. \ref{fig:aloha_vs_cdma}, we evaluated the performance for a short-read time set to $128$ slots for the slotted ALOHA and spreading code length of $128$ for the CDMA scheme. The spread signal from a code length of $128$ will take the same transmit duration as $128$ slots in the ALOHA. Results show the superior performance of CDMA for all numbers of motes beyond $20$, with a decline in performance at $120$ motes. This is due to ALOHA's low utilization efficiency of the $128$ available slots, as most slots are not selected in the random back-off process and remain empty. For long-read times, the analysis in Fig. \ref{fig:aloha_vs_cdma2} shows a superior performance of the slotted ALOHA scheme for number of deployed motes up to $50$, beyond which performance begins to oscillate around that of the CDMA. For a much longer read duration, by either increasing the communication rate or the number of available slots, the ALOHA scheme will read all the deployed motes. The increase in duration has no impact on the CDMA performance. We can infer from the results that for implantable medical device applications that require a short-read time, CDMA will be a better choice but at the expense of implementing more complex procedures in micro-scale.  Slotted ALOHA can be used in applications where long-read durations can be tolerated.



\section{Conclusion}\label{sec:Conclusion}
In this work, we first established the major challenges of implementing the existing wireless communication technologies in micro-scale implantable medical devices which we referred to as ``biomotes''. To overcome these challenges, we proposed the use of magnetic induction-based backscatter communication. First, we demonstrated that communication via magnetic induction can be established between a micro-scale IMD and a handheld device as an external reader. To increase the communication range, we adapted suitable low-power modulation and error-correction coding schemes at the physical layer. We also investigated the feasibility of implementing these schemes in micro-scale. To increase the reliability and accuracy of measurements via mass deployment, we proposed suitable low-power MAC schemes and investigated the feasibility of their implementation in micro-scale. Assuming reasonable design parameters and the human body as an AWGN channel, performance of the proposed biomote was analyzed. For a BER of $10^{-3}$ both with and without error-correction coding, we obtained communication ranges of between $5$ and $6$ cm for BPSK and ASK modulations. Results obtained from our analysis of the MAC schemes can be used as guidelines for selecting mote design parameters such as data rate, as well as mote utilization specifications such as read duration required and optimum number of motes to be deployed. While mass deployment of such IMDs can increase reliability, MAC schemes with simple complexity must be incorporated. In summary, it was found that magnetic induction is a formidable candidate for communication between micro-scale IMDs and external readers such as NFC-enabled handheld devices.  

\bibliographystyle{ieeetran}
\bibliography{Ref_new}

\begin{thebibliography}{10}
\providecommand{\url}[1]{#1}
\csname url@samestyle\endcsname
\providecommand{\newblock}{\relax}
\providecommand{\bibinfo}[2]{#2}
\providecommand{\BIBentrySTDinterwordspacing}{\spaceskip=0pt\relax}
\providecommand{\BIBentryALTinterwordstretchfactor}{4}
\providecommand{\BIBentryALTinterwordspacing}{\spaceskip=\fontdimen2\font plus
\BIBentryALTinterwordstretchfactor\fontdimen3\font minus
  \fontdimen4\font\relax}
\providecommand{\BIBforeignlanguage}[2]{{%
\expandafter\ifx\csname l@#1\endcsname\relax
\typeout{** WARNING: IEEEtran.bst: No hyphenation pattern has been}%
\typeout{** loaded for the language `#1'. Using the pattern for}%
\typeout{** the default language instead.}%
\else
\language=\csname l@#1\endcsname
\fi
#2}}
\providecommand{\BIBdecl}{\relax}
\BIBdecl

\bibitem{vuka2017mobile}
P.~Vuka, A.~Behfarnia, and A.~Eslami, ``A mobile-enabled micro communication
  device for biosensing,'' in \emph{17th International Conference on
  Nanotechnology (IEEE-NANO)}, Pittsburgh, PA, USA, July 2017, pp. 65--68.

\bibitem{durairaj2012nano}
R.~Durairaj, J.~Shanker, and M.~Sivasankar, ``Nano robots in bio medical
  application,'' in \emph{IEEE International Conference on Advances in
  Engineering, Science and Management (ICAESM)}, Nagapattinam, Tamil Nadu,
  India, March 2012, pp. 67--72.

\bibitem{bourzac2012carrying}
K.~Bourzac, ``Carrying drugs,'' \emph{Nature}, vol. 491, no. 7425, pp.
  S58--S60, 2012.

\bibitem{yates2013life}
M.~R. Yates and C.~Y. Barlow, ``Life cycle assessments of biodegradable,
  commercial biopolymers{: A} critical review,'' \emph{Resources, Conservation
  and Recycling}, vol.~78, pp. 54--66, 2013.

\bibitem{hong2018compact}
S.~Hong, W.~Jung, T.~Kim, and K.~Oh, ``Compact biocompatible fiber optic
  temperature microprobe using {DNA}-based biopolymer,'' \emph{Journal of
  Lightwave Technology}, vol.~36, no.~4, pp. 974--978, 2018.

\bibitem{haselman2010future}
M.~Haselman and S.~Hauck, ``The future of integrated circuits: A survey of
  nanoelectronics,'' \emph{Proceedings of the IEEE}, vol.~98, no.~1, pp.
  11--38, January 2010.

\bibitem{liu2008mems}
J.-Q. Liu, H.-B. Fang, Z.-Y. Xu, X.-H. Mao, X.-C. Shen, D.~Chen, H.~Liao, and
  B.-C. Cai, ``A {MEMS}-based piezoelectric power generator array for vibration
  energy harvesting,'' \emph{Microelectronics Journal}, vol.~39, no.~5, pp.
  802--806, 2008.

\bibitem{Deterre2012}
M.~Deterre, E.~Lefeuvre, and E.~Dufour-Gergam, ``An active piezoelectric energy
  extraction method for pressure energy harvesting,'' \emph{Smart Materials and
  Structures}, vol.~21, no.~8, p. 085004, 2012.

\bibitem{mujeeb2015optical}
M.~Mujeeb-U-Rahman, D.~Adalian, C.-F. Chang, and A.~Scherer, ``Optical power
  transfer and communication methods for wireless implantable sensing
  platforms,'' \emph{Journal of Biomedical Optics}, vol.~20, no.~9, pp.
  095\,012--095\,012, 2015.

\bibitem{Charthad2015mm}
J.~Charthad, M.~J. Weber, T.~C. Chang, and A.~Arbabian, ``A mm-sized
  implantable medical device ({IMD}) with ultrasonic power transfer and a
  hybrid bi-directional data link,'' \emph{IEEE Journal of Solid-State
  Circuits}, vol.~50, no.~8, pp. 1741--1753, 2015.

\bibitem{Ho2014}
J.~S. Ho, A.~J. Yeh, E.~Neofytou, S.~Kim, Y.~Tanabe, B.~Patlolla, R.~E. Beygui,
  and A.~S. Poon, ``Wireless power transfer to deep-tissue microimplants,''
  \emph{Proceedings of the National Academy of Sciences}, vol. 111, no.~22, pp.
  7974--7979, 2014.

\bibitem{jornet2011channel}
J.~M. Jornet and I.~F. Akyildiz, ``Channel modeling and capacity analysis for
  electromagnetic wireless nanonetworks in the terahertz band,'' \emph{IEEE
  Transactions on Wireless Communications}, vol.~10, no.~10, pp. 3211--3221,
  2011.

\bibitem{akyildiz2016realizing}
I.~F. Akyildiz and J.~M. Jornet, ``Realizing ultra-massive mimo (1024$\times$
  1024) communication in the (0.06--10) terahertz band,'' \emph{Nano
  Communication Networks}, vol.~8, pp. 46--54, 2016.

\bibitem{akyildiz2014teranets}
I.~Akyildiz, J.~Jornet, and C.~Han, ``Teranets: Ultra-broadband communication
  networks in the terahertz band,'' \emph{IEEE Wireless Communications},
  vol.~21, no.~4, pp. 130--135, 2014.

\bibitem{akkari2016joint}
N.~Akkari, J.~M. Jornet, P.~Wang, E.~Fadel, L.~Elrefaei, M.~G.~A. Malik,
  S.~Almasri, and I.~F. Akyildiz, ``Joint physical and link layer error control
  analysis for nanonetworks in the terahertz band,'' \emph{Wireless Networks},
  vol.~22, no.~4, pp. 1221--1233, 2016.

\bibitem{Peisino2013}
M.~Peisino and P.~Ryser, ``Deeply implanted medical device based on a novel
  ultrasonic telemetry technology,'' Ph.D. dissertation, Ecole Polytechnique
  Federale de Lausanne, Switzerland, 2013.

\bibitem{Davilis2010}
Y.~Davilis, A.~Kalis, and A.~Ifantis, ``On the use of ultrasonic waves as a
  communications medium in biosensor networks,'' \emph{IEEE Transactions on
  Information Technology in Biomedicine}, vol.~14, no.~3, pp. 650--656, 2010.

\bibitem{Galluccio2012challenges}
L.~Galluccio, T.~Melodia, S.~Palazzo, and G.~E. Santagati, ``Challenges and
  implications of using ultrasonic communications in intra-body area
  networks,'' in \emph{9th Annual Conference on Wireless On-demand Network
  Systems and Services (WONS)}, Courmayeur, Italy, January 2012, pp. 182--189.

\bibitem{schuettler2012hermetic}
M.~Schuettler, F.~Kohler, J.~S. Ordonez, and T.~Stieglitz, ``Hermetic
  electronic packaging of an implantable brain-machine-interface with
  transcutaneous optical data communication,'' in \emph{IEEE Annual
  International Conference of the Engineering in medicine and Biology Society
  (EMBC)}, San Diego, CA, USA, September 2012, pp. 3886--3889.

\bibitem{nafari2015metallic}
M.~Nafari and J.~M. Jornet, ``Metallic plasmonic nano-antenna for wireless
  optical communication in intra-body nanonetworks,'' in \emph{Proceedings of
  the 10th EAI International Conference on Body Area Networks}, Sydney, New
  South Wales, Australia, September 2015, pp. 287--293.

\bibitem{song2012signal}
Y.~Song, Q.~Hao, K.~Zhang, J.~Wang, X.~Jin, and H.~Sun, ``Signal transmission
  in a human body medium-based body sensor network using a {Mach-Zehnder}
  electro-optical sensor,'' \emph{Sensors}, vol.~12, no.~12, pp.
  16\,557--16\,570, 2012.

\bibitem{shinagawa2013compact}
M.~Shinagawa, A.-i. Sasaki, A.~Furuya, H.~Morimura, and K.~Aihara, ``Compact
  electro-optic sensor module for intra-body communication using optical pickup
  technology,'' \emph{Japanese Journal of Applied Physics}, vol.~52, no. 9S2,
  p. 09LA03, 2013.

\bibitem{zakrajsek2016lithographically}
L.~Zakrajsek, E.~Einarsson, N.~Thawdar, M.~Medley, and J.~M. Jornet,
  ``Lithographically defined plasmonic graphene antennas for terahertz-band
  communication,'' \emph{IEEE Antennas and Wireless Propagation Letters},
  vol.~15, pp. 1553--1556, 2016.

\bibitem{jornet2013graphene}
J.~M. Jornet and I.~F. Akyildiz, ``Graphene-based plasmonic nano-antenna for
  terahertz band communication in nanonetworks,'' \emph{IEEE Journal on
  Selected Areas in Communications}, vol.~31, no.~12, pp. 685--694, 2013.

\bibitem{llatser2012graphene}
I.~Llatser, C.~Kremers, A.~Cabellos-Aparicio, J.~M. Jornet, E.~Alarc{\'o}n, and
  D.~N. Chigrin, ``Graphene-based nano-patch antenna for terahertz radiation,''
  \emph{Photonics and Nanostructures-Fundamentals and Applications}, vol.~10,
  no.~4, pp. 353--358, 2012.

\bibitem{pickwell2006biomedical}
E.~Pickwell and V.~Wallace, ``Biomedical applications of terahertz
  technology,'' \emph{Journal of Physics D: Applied Physics}, vol.~39, no.~17,
  p. R301, 2006.

\bibitem{oh2013measurement}
S.~J. Oh, S.-H. Kim, K.~Jeong, Y.~Park, Y.-M. Huh, J.-H. Son, and J.-S. Suh,
  ``Measurement depth enhancement in terahertz imaging of biological tissues,''
  \emph{Optics Express}, vol.~21, no.~18, pp. 21\,299--21\,305, 2013.

\bibitem{federal1996guidelines}
F.~C. Commission \emph{et~al.}, ``Guidelines for evaluating the environmental
  effects of radiofrequency radiation,'' \emph{Report and Order, ET Docket
  93-62, FCC 96-326}, 1996.

\bibitem{ieee1992ieee}
I.~S.~C. Committee \emph{et~al.}, ``{IEEE} standard for safety levels with
  respect to human exposure to radio frequency electromagnetic fields, 3khz to
  300ghz,'' \emph{IEEE C95. 1-1991}, 1992.

\bibitem{freitas1999nanomedicine}
R.~A. Freitas, \emph{Nanomedicine, volume I: {B}asic capabilities}.\hskip 1em
  plus 0.5em minus 0.4em\relax Landes Bioscience, Georgetown, TX, 1999, vol.~1.

\bibitem{dove2003notes}
E.~L. Dove, ``{Notes on ultrasound-echocardiography 51: 060 fundamentals of
  bioimaging},'' \emph{Ultrasound Dove}, 2003.

\bibitem{cannata2003design}
J.~M. Cannata, T.~A. Ritter, W.-H. Chen, R.~H. Silverman, and K.~K. Shung,
  ``Design of efficient, broadband single-element (20-80 mhz) ultrasonic
  transducers for medical imaging applications,'' \emph{IEEE Transactions on
  Ultrasonics, Ferroelectrics, and Frequency Control}, vol.~50, no.~11, pp.
  1548--1557, 2003.

\bibitem{pierobon2014routing}
M.~Pierobon, J.~M. Jornet, N.~Akkari, S.~Almasri, and I.~F. Akyildiz, ``A
  routing framework for energy harvesting wireless nanosensor networks in the
  terahertz band,'' \emph{Wireless Networks}, vol.~20, no.~5, pp. 1169--1183,
  2014.

\bibitem{akyildiz2010electromagnetic}
I.~F. Akyildiz and J.~M. Jornet, ``Electromagnetic wireless nanosensor
  networks,'' \emph{Nano Communication Networks}, vol.~1, no.~1, pp. 3--19,
  2010.

\bibitem{Seo2013}
D.~Seo, J.~M. Carmena, J.~M. Rabaey, E.~Alon, and M.~M. Maharbiz, ``Neural
  dust: An ultrasonic, low power solution for chronic brain-machine
  interfaces,'' \emph{arXiv preprint arXiv:1307.2196}, 2013.

\bibitem{ghovanloo2007wide}
M.~Ghovanloo and S.~Atluri, ``A wide-band power-efficient inductive wireless
  link for implantable microelectronic devices using multiple carriers,''
  \emph{IEEE Transactions on Circuits and Systems I: Regular Papers}, vol.~54,
  no.~10, pp. 2211--2221, 2007.

\bibitem{sutardja2017isolator}
C.~Sutardja and J.~Rabaey, ``Isolator-less near-field rfid reader for
  sub-cranial powering/data link of mm-sized implants,'' in \emph{43rd IEEE
  European Solid State Circuits Conference (ESSCIRC)}, Leuven, Belgium,
  September 2017, pp. 372--375.

\bibitem{Goodarzy2015}
F.~Goodarzy, E.~. Skafidas, and S.~Gambini, ``Feasibility of energy-autonomous
  wireless microsensors for biomedical applications: Powering and
  communication,'' \emph{IEEE Reviews in Biomedical Engineering}, vol.~8, pp.
  17--29, 2015.

\bibitem{Simard2010}
G.~Simard, M.~Sawan, and D.~Massicotte, ``High-speed {OQPSK} and efficient
  power transfer through inductive link for biomedical implants,'' \emph{IEEE
  Transactions on Biomedical Circuits and Systems}, vol.~4, no.~3, pp.
  192--200, 2010.

\bibitem{wiki_phone}
W.~contributors, ``Iphone --- {Wikipedia}{,} the free encyclopedia,''
  \url{https://en.wikipedia.org/w/index.php?title=IPhone&oldid=840824257},
  2018, [Online; accessed 16-May-2018].

\bibitem{sun2010magnetic}
Z.~Sun and I.~F. Akyildiz, ``Magnetic induction communications for wireless
  underground sensor networks,'' \emph{IEEE Transactions on Antennas and
  Propagation}, vol.~58, no.~7, pp. 2426--2435, 2010.

\bibitem{3DCoil97}
W.~M. Ng, C.~Zhang, D.~Lin, and S.~R. Hui, ``Two-and three-dimensional
  omnidirectional wireless power transfer,'' \emph{IEEE Transactions on Power
  Electronics}, vol.~29, no.~9, pp. 4470--4474, 2014.

\bibitem{3DCoil14}
H.~Guo and Z.~Sun, ``Channel and energy modeling for self-contained wireless
  sensor networks in oil reservoirs,'' \emph{IEEE Transactions on Wireless
  Communications}, vol.~13, no.~4, pp. 2258--2269, 2014.

\bibitem{hu2005science}
J.~Hu and M.~Yan, ``Science letters: Preparation of high-permeability nicuzn
  ferrite,'' \emph{Journal of Zhejiang University. Science. B}, vol.~6, no.~6,
  p. 580, 2005.

\bibitem{LTE_RSRP}
G.~T. 36.133, ``Evolved universal terrestrial radio access ({E-UTRA});
  {Requirements} for support of radio resource management.''

\bibitem{LTE_RSRP2}
Aroccasolutions, ``{RSRP and RSRQ Measurement in LTE},''
  \url{https://www.laroccasolutions.com/78-rsrp-and-rsrq-measurement-in-lte},
  2016, [Online; accessed 16-May-2018].

\bibitem{cho2006ask}
J.~Cho, K.-W. Min, and S.~Kim, ``An {ASK} modulator and antenna driver for
  13.56 {MHz RFID readers and NFC devices},'' \emph{IEICE Transactions on
  Communications}, vol.~89, no.~2, pp. 598--600, 2006.

\bibitem{Hassouni2015design}
S.~Hassouni and H.~Qjidaa, ``A design of modulator and demodulator for a
  passive {UHF RFID tag using DTMOST compatible with C1 G2 EPC} standard
  protocol,'' \emph{International Journal of Wireless Information Networks},
  vol.~22, no.~4, pp. 407--414, 2015.

\bibitem{burasa2016high}
P.~Burasa, T.~Djerafi, N.~G. Constantin, and K.~Wu, ``High-data-rate
  single-chip battery-free active millimeter-wave identification tag in 65-nm
  cmos technology,'' \emph{IEEE Transactions on Microwave Theory and
  Techniques}, vol.~64, no.~7, pp. 2294--2303, 2016.

\bibitem{yang2012product}
C.~Yang, Y.~Emre, and C.~Chakrabarti, ``Product code schemes for error
  correction in {MLC NAND} flash memories,'' \emph{IEEE Transactions on Very
  Large Scale Integration (VLSI) Systems}, vol.~20, no.~12, pp. 2302--2314,
  2012.

\bibitem{bletsas2009anti}
A.~Bletsas, S.~Siachalou, and J.~N. Sahalos, ``Anti-collision backscatter
  sensor networks,'' \emph{IEEE Transactions on Wireless Communications},
  vol.~8, no.~10, 2009.

\bibitem{das2014performance}
S.~Das, I.~Banerjee, M.~Chatterjee, and T.~Samanta, ``Performance analysis of
  tdma based data transmission in wsn,'' in \emph{14th International Conference
  on Intelligent Systems Design and Applications (ISDA)}, Okinawa, Japan,
  November 2014, pp. 107--112.

\bibitem{salles2017dissipated}
V.~Salles, S.~Barbin, and L.~Kretly, ``Dissipated energy of a low-power
  adiabatic {CPAL JK-FF design using four-phase AC-clocked power supply based
  on 180 nm CMOS} technology with various load capacitances,'' in \emph{IEEE
  APS Topical Conference on Antennas and Propagation in Wireless Communications
  (APWC)}, Verona, Italy, September 2017, pp. 1933--1935.

\bibitem{zhang2014low}
M.~Zhang, L.~Cai, X.~Yang, H.~Cui, Z.~Wang, C.~Feng, and S.~Wang, ``Low power
  on-chip clocking for nanomagnetic logic circuits,'' \emph{Micro \& Nano
  Letters}, vol.~9, no.~10, pp. 753--755, 2014.

\bibitem{Datta2017}
D.~Datta, B.~Datta, and H.~S. Dutta, ``Design and implementation of multibit
  {LFSR on FPGA} to generate pseudorandom sequence number,'' in \emph{Devices
  for Integrated Circuit (DevIC)}, 2017, pp. 346--349.

\bibitem{bharti2017implementation}
A.~Bharti and A.~Sharma, ``Implementation of {Ex-OR gate using QCA with NNI
  logic},'' in \emph{IEEE International Conference on Wireless Communications,
  Signal Processing and Networking (WiSPNET)}, Chennai, India, March 2017, pp.
  2027--2031.

\bibitem{Cox2005time}
D.~Cox, E.~Jovanov, and A.~Milenkovic, ``Time synchronization for zigbee
  networks,'' in \emph{37th Southeastern Symposium on System Theory (SSST)},
  Tuskegee, AL, USA, March 2005, pp. 135--138.

\bibitem{Luby1996}
M.~G. Luby and M.~Luby, \emph{Pseudorandomness and cryptographic
  applications}.\hskip 1em plus 0.5em minus 0.4em\relax Princeton University
  Press, 1996.

\bibitem{mandal2016}
K.~Mandal, X.~Fan, and G.~Gong, ``Design and implementation of warbler family
  of lightweight pseudorandom number generators for smart devices,'' \emph{ACM
  Transactions on Embedded Computing Systems (TECS)}, vol.~15, no.~1, p.~1,
  2016.

\bibitem{benkic2007proposed}
K.~Benkic, ``Proposed use of a {CDMA} technique in wireless sensor networks,''
  in \emph{14th International Workshop on Systems, Signals and Image Processing
  and 6th EURASIP Conference focused on Speech and Image Processing, Multimedia
  Communications and Services}, Maribor, Slovenia, June 2007, pp. 343--348.

\bibitem{wiki_noise}
W.~contributors, ``Johnson --- nyquist noise,{Wikipedia}{,} the free
  encyclopedia,''
  \url{https://en.wikipedia.org/w/index.php?title=Johnson%E2%80%93Nyquist_noise&oldid=837565746},
  2018, [Online; accessed 16-May-2018].

\bibitem{Noise_Fig}
A.~Kumar, W.~A. Edelstein, and P.~A. Bottomley, ``Noise figure limits for
  circular loop {MR} coils,'' \emph{Magnetic Resonance in Medicine}, vol.~61,
  no.~5, pp. 1201--1209, 2009.

\end{thebibliography}
\IEEEpeerreviewmaketitle

\end{document}